\documentclass[british, 12pt, a4paper]{article}
\usepackage[left=1in,top=1in,right=1in,bottom=1in,nohead,paperwidth=8.5in, paperheight=11in]{geometry} %1 inch margins, 8 1/2 x 11-inch pages: standard choices for most journals
\usepackage{verbatim,color,setspace,graphicx,epsfig,rotating, epstopdf,lscape}
\usepackage{amsmath,amsfonts, lineno, hyperref, geometry, bbm} 
\usepackage{arydshln}
\usepackage[normalem]{ulem}
\usepackage[utf8]{inputenc}
\usepackage[british]{babel}
\usepackage{comment}
\usepackage[round]{natbib} %standard package for references
\usepackage[title]{appendix}
\usepackage{xcolor}
\usepackage{setspace}

\begin{document}

\def\spacingset#1{\renewcommand{\baselinestretch}%
{#1}\small\normalsize} \spacingset{1}

\title{
Forecasting GDP in Europe with Textual Data
}
\author{Luca Barbaglia$^{a}$, Sergio Consoli$^{a}$, Sebastiano Manzan$^{b}$
\\
\mbox{}\\
{\small $^{a}$ European Commission, Joint Research Centre (JRC), Ispra (VA), Italy}\\
{\small $^{b}$ Zicklin School of Business, Baruch College, New York 10010, USA}\\
\\ \mbox{} \\
%{PRELIMINARY: DO NOT CITE}
}

\date{\today}
%\date{}

\date{}
\maketitle
\thispagestyle{empty}
\setcounter{page}{1}

% \vspace{-0.5cm}
% \begin{center}
% \Large{\textsc{Preliminary and incomplete: \\ please do not circulate}}
% \end{center}

\vspace{0.1cm}
\begin{abstract}
\begin{center}
\mbox{}\\
\begin{minipage}{13cm}
\noindent
{\small 
We evaluate the informational content of news-based sentiment indicators for forecasting Gross Domestic Product (GDP) {\color{black} and other macroeconomic variables} of the five major European economies.  Our data set includes over 27 million articles for 26 major newspapers in 5 different languages. The evidence indicates that these sentiment indicators are significant predictors to forecast macroeconomic variables and their predictive content is robust to controlling for other indicators available to forecasters in real-time. 
} 
%\mbox{}\\
\newline\mbox{}
\\
\end{minipage}
\end{center}

\end{abstract}

\vspace{0.4cm}
\begin{minipage}{15cm}
\singlespacing
{\footnotesize 
The views expressed are purely those of the authors and do not, in any circumstance, be regarded as stating an official position of the European Commission. 
We are grateful to the participants of the 41$^{st}$ International Symposium on Forecasting,
the Big Data \& Economic Forecasting workshop held at the European Commission, 
the Real-time data analysis, methods and application conference held at the Banque de France,
seminar participants at the Bank of Latvia and at the National Bank of Belgium, {\color{black} and the Editor and reviewers} for numerous comments that significantly improved the paper. 
We are greatly indebted to the Centre for Advanced Studies at the Joint Research Centre of the European Commission for the support, encouragement, and stimulating environment while working on the {\it bigNOMICS} project. 
The authors thank Susan Wang for the support with the topic modeling tasks.
{\tt E-mail}: luca.barbaglia@ec.europa.eu, sergio.consoli@ec.europa.eu, and sebastiano.manzan@baruch.cuny.edu. }
\end{minipage}

\newpage
\doublespacing

\section{Introduction}
\label{sec:introduction}

%% SURVEYS and PROBLEM
Business and consumer surveys are an essential tool used by policy-makers and practitioners to monitor and forecast the economy.
%\citep{christiansen2014forecasting, wilms2016predictive, lehmann2020forecasting}. 
Their most valuable feature is to provide timely information about the current and expected state of economic activity that is relevant to integrate the sluggish release of macroeconomic indicators. 
Interestingly, surveys are often interpreted as measures of economic sentiment in the sense of providing the pulse of different aspects of the economy, such as the consumers' attitude toward spending or the expectation of purchasing managers about inflation. 
Some prominent examples are represented by the Survey of Consumers of the University of Michigan (MCS) for the United States \citep{curtin2015university} and the Business and Consumer Survey (BCS) for the European Union %\footnote{More information is available at \url{https://ec.europa.eu/info/business-economy-euro/indicators-statistics/economic-databases/business-and-consumer-surveys_en}.}
\citep{ec_survey}. 
Although surveys are very valuable and accurate proxies of economic activity, they are {\color{black} typically} released  at the monthly frequency {\color{black} which might limit their usefulness in high-frequency nowcasting of economic variables} \citep{aguilar2020can, algaba2020daily}. 
%Another issue is that surveys are designed to measure sentiment regarding specific topics of interest that cannot very easily be altered.
% However, their usefulness has been debated in the recent literature, with several findings that surveys have predictive power for economic forecasting (\citealp{christiansen2014forecasting} among others), while others indicating that these marginal gains are rarely significant \citep{claveria2007business} or relevant only for specific industries \citep{wilms2016predictive}.
% These indicators are the product of well-established and tested sampling strategies: they collect the opinion of economic agents and provide a proxy of the current and future tendency of a specific aspect of the economy \citep{lehmann2020forecasting}.

%% FIGAS SENTIMENT
The goal of this paper is to contribute an alternative measure of economic sentiment that, similarly to surveys, captures the overall attitude of the public toward specific aspects of the economy. The measure that we propose is obtained from textual analysis of a large data set of daily newspaper articles. We believe that news represent a valuable source of information as they report on economic and political events that could possibly influence economic decisions. In this sense, we expect that measuring economic sentiment from news text can provide a signal regarding the current state of the economy as well as its future trajectory. 
%Our objective is to create indicators of sentiment that, similarly to surveys, refer to specific aspects of interest, such as the industry or services confidence indexes in the case of the BCS. 
% \sout{We achieve this by isolating the text in the news articles that refers to a specific topic of interest and perform sentiment analysis only on the relevant piece of news. 
% Then, we analyze the dependence structure within the text to determine the words that are semantically related to the topic of interest: the aim is to accurately identify only the text that closely refers to and characterizes the topic of interest (e.g., economy or manufacturing), and compute the \textit{aspect-based} sentiment only on those terms.
% Finally, we assign a \textit{fine-grained} sentiment score to the selected words represented by a numerical value in the range [-1, 1]. 
% This allows adapting the sentiment score to the emotional content of the text, as opposed to classifying words in the positive and negative categories. }
%These three features characterize the Fine-Grained Aspect-based Sentiment (FiGAS) approach that we use to compute our sentiment indicators \cite[see][]{Barbaglia2020101,barbaglia2020forecasting}. 
%With respect to surveys, the proposed indicators are available timely at daily frequencies and the allow to flexibly select the topic about which one shall compute the sentiment. 
%% ADVANTAGES OF FIGAS
Deriving sentiment measures from news provides several advantages.
One is that sentiment can be measured at high frequencies since news are available on a daily basis, as opposed to surveys that are typically available at the monthly frequency as discussed earlier. 
Such a feature can be very helpful when the economy is rapidly changing direction and high-frequency monitoring in real-time is needed. 
Another convenient feature of using news is that they allow to compute sentiment about any economic aspect of interest at a small additional cost relative to surveys that have a fixed structure and cannot be changed very easily. 
For instance, the MCS and BCS are mostly focused on surveying the attitude of consumers and businesses regarding employment, output, and prices, while neglecting topics such as financial conditions and monetary policy. 
Overall, the findings in \cite{larsen2021news} show that news have a significant role in shaping the inflation expectations of consumers.  A relevant question to answer is whether there is any residual predictability in the news-based sentiment {\color{black} once we take into account the sentiment from surveys}. 
%Overall, it is likely that sentiment derived from surveys provides a more accurate signal relative to the news-based one. 
%{\color{purple}This is because surveys reflect the attitudes, expectations, and purchasing intentions of economic agents as opposed to news that report on a wide range of economic events.}
%{\color{black} News also expectations are not easy to disentangle from the reporting of past events}
%However, we expect news-based sentiment to provide a timely, although probably noisier, measure of the state of the economy relative to the survey's confidence indicators and macroeconomic variables. 
In this paper, we thus focus on understanding the incremental value of sentiment constructed from textual news data in the context of {\color{black} nowcasting and } 
 forecasting GDP for five European countries that are characterized by considerable delays in the release of their official statistics. 
%% DELAYS https://ec.europa.eu/eurostat/statistics-explained/index.php?title=Preliminary_GDP_flash_estimate_in_30_days_for_Europe
\textcolor{black}{
{\color{black} Estimating GDP is a very complex task that requires to gather information about the economic activities of thousands of consumers and businesses mostly via surveys. The data collection process is quite involved and time consuming which leads to significant delays in releasing macroeconomic data}. For instance, Eurostat  publishes the  preliminary flash estimate of GDP 30 days after the end of the quarter {\color{black} based on one or at best two releases of the monthly indicators.} {\color{black} More accurate estimates are released with the} official flash estimate  {\color{black} after 45 days and the} traditional (non-flash) estimates are normally published with a 65 days delay \citep{eurostat2016euro}.
}
%% WHY ONLY GDP?

%% THIS PAPER
We construct the sentiment indicators based on a data set that includes news from 26 major newspapers in France, Germany, Italy, Spain and the UK. The data set amounts to over 27 million articles and a total of 12.5 billion words in 5 languages. 
To transform the text into a sentiment measure we follow the Fine-Grained, Aspect-based Sentiment (FiGAS) approach proposed by \citet{consoli2021finegrained} and applied to the US by \citet{barbaglia2020forecasting}. 
% In the first step of the analysis we isolate the sentences in the news articles that refer to a specific topic of interest (e.g., economy or unemployment). Then, we analyze the dependence structure within the sentence to determine the words that are linguistically related to the aspect of interest. The aim is to accurately identify the words in the sentence that characterize the topic of interest and compute sentiment only on those terms. The final step consists of assigning a fine-grained sentiment score between -1 and 1 to the selected words. 
% %This allows adapting the sentiment score to the intensity of the emotional content of the text, as opposed to simply classifying words in the positive and negative categories.
% This allows adapting the sentiment score to the intensity of the emotional content about the topic of interest, as opposed to simply classifying words in the positive and negative categories.
The FiGAS approach has been originally designed for the analysis of text in the English language, and thus cannot be easily adapted to the analysis of text in other languages. To overcome this problem, we translate the articles from French, German, Spanish and Italian to English relying on a neural machine %automatic 
translation service, which contributes to further increasing the computational burden of the analysis. 
% This work contributes to an emerging literature that incorporates sentiment from news, where many authors have investigated their added-value for forecasting economic growth while considering single countries, like the US \citep{ardia2019questioning, barbaglia2020forecasting, ellingsen2020news, shapiro2020measuring}, the UK \citep{rambaccussing2020forecasting, kalamaramaking}, France \citep{bortoli2018nowcasting}, Germany \citep{feuerriegel2019news}, Italy \citep{aprigliano2021power}, Norway \citep{larsen2019value} or Spain \citep{aguilar2020can}.
This work joins %contributes 
 an emerging literature that uses textual data from news to forecast economic variables with a focus, in particular, on the US \citep{ardia2019questioning, barbaglia2020forecasting, ellingsen2020news, SHAPIRO2022221}, the UK \citep{rambaccussing2020forecasting, kalamaramaking}, France \citep{bortoli2018nowcasting}, Germany \citep{feuerriegel2019news}, Italy \citep{aprigliano2021power}, Norway \citep{larsen2019value}, and Spain \citep{aguilar2020can}.
There are very few studies in economics and finance that perform a textual analysis of news across several countries and languages \citep[e.g., ][]{baker2016measuring, ashwin2021nowcasting}.
Although this adds considerable complications, in particular in the computational aspect, we believe that it offers the opportunity to validate the robustness of our findings across countries and to reveal country-specific characteristics of the relationship between news and macroeconomic variables.
%effects of news on macroeconomic variables.
%This work contributes to an emerging literature on incorporating sentiment from news in forecasting models while employing cutting-edge NLP techniques in a multi-language and multi-country framework.

%% RESULTS
We design six sentiment indicators to capture the attitude in the news regarding the economy, unemployment rate, inflation, manufacturing, financial sector, and monetary policy. The choice of a wide range of topics aims at capturing various aspects of economic activity and policy.
We find that the sentiment measures related to the real sector have a significant business cycle component in all countries considered. 
In addition, the evidence indicates that these measures are highly correlated across countries, and in particular in the case of the {\it Monetary Policy} and {\it Inflation} sentiments for the countries in the euro area. 
This is consistent with the fact that these countries participate in a monetary union that contributes to synchronising news about monetary policy, such as 
the actions of the European Central Bank.

% Adopting a Granger-causal perspective, we show that news-based indicators provide a timely approximation of official surveys, with the advantage of covering a diverse set of topics and more broadly represent the various aspect of economic activity.
 {\color{black}  The results indicate also } that the sentiment measures are significant predictors for GDP growth at horizons ranging from 30 days to a year before the official release. 
This result is robust to controlling for the real-time flow of macroeconomic releases as well as survey information. 
In this sense, we believe that the predictive power of sentiment is genuinely incremental relative to the informational content of survey confidence indexes.
Furthermore, we find that the sentiment indicators that are relevant to predict GDP are heterogeneous across countries. 
In the case of Germany, the predictive content is mostly embedded in the news regarding the unemployment rate and inflation, while news on monetary policy appear to be particularly relevant in Spain and Italy. 
We also find that news discussing the financial sector are informative in the case of France and the UK. 
These results suggest that the attention of the public to news might depend on country-specific characteristics, such as the Germans' concern %preoccupation 
with inflation or the Italians' sensitivity to monetary policy decisions. {\color{black} In addition, we also validate our findings for GDP on other macroeconomic variables of interest, such as the unemployment  and inflation rates, and the growth rate of industrial production and confirm, to a large extent, the evidence that sentiment measures provide genuine predictive power.}

The paper is structured as follows. 
Section \ref{sec_methodology} introduces the sentiment-based indicators and the forecasting methodology, while Section \ref{sec_data} describes the data and discusses the proposed news-based indicators and their relation with official surveys.
The in-sample and out-of-sample real-time forecasting exercises are carried out in Sections \ref{sec_insample} and \ref{sec_oos}, respectively, while Section \ref{sec:other_macrovariables} {\color{black} evaluates the robustness of our findings to forecast other macroeconomic variables.}
Finally, Section \ref{sec_conclusion} concludes the paper.% and considers some directions for future research.

\section{Methodology\label{sec_methodology}}
We follow the FiGAS approach proposed by \citet{consoli2021finegrained} to construct sentiment indicators of different aspects of economic activity at the daily frequency. A key feature of FiGAS is that it computes sentiment that relates to an economic topic (e.g., {\it Economy} or {\it Financial Sector}) by considering only the words that are connected to the terms of interest in the sentence. %article. 
More specifically, the goal is to isolate the neighbouring words that are grammatically related to the concept %term(s) 
of interest and that modify and characterize its tone. %Indeed, not all words neighbouring a term of interest are considered to carry sentiment, but only those that modify and characterize its connotation. 
%modify its meaning. 
The approach produces a sentiment indicator that is targeted to monitor the attitude in the news regarding a topic of interest, as opposed to alternative approaches that calculate sentiment for the entire article. 
%An indicator at the article level 
Such an approach might be able to capture the sentiment regarding a wide range of concepts that are discussed in the news text and could be useful to measure the overall state of the economy. 
%However, pooling the text in an article might not measure accurately the sentiment towards a specific economic aspect that we might be interested to monitor. 
In this sense, the FiGAS sentiment is \textit{aspect-based} since it refers to a particular topic. 
This could be relevant for policy-makers when the concern is to extract the sentiment regarding a specific variable, for example inflation, rather than an aggregate indicator.

% In addition, the approach is {\it fine-grained} since it relies on a dictionary that assigns a score to each word in the range $[-1, 1]$, rather than simply classifying words in positive and negative. 
% Overall, the FiGAS approach promises to deliver a sentiment that is targeted to the topic of interest and accurate given its fine-grained nature. 
% Another characteristic of this approach is that it aims at extracting from the sentence and the article the reference to a geographic location. 
% The goal is to include in the calculation of the sentiment only the text that references to the country we are targeting to monitor. 
% In case no country can be identified in the text we assume that it refers to the place where the journal is published. 
% We refer to \citep{consoli2021finegrained} for a detailed presentation of the algorithm.

%% FINE-GRAINED
%Once we have identified pieces of text containing those terms, the FiGAS analysis continues by parsing its semantic dependence in order to isolate the words that are related to the identified term of interest. 
The sentiment of the text is then obtained by assigning a category to each word (e.g., positive or negative) or a numerical value that measures the sentiment content of the word. 
Our approach is {\it fine-grained} since it relies on a dictionary that scores words in the range $[-1, 1]$, rather than simply classifying them in predefined categories \citep{loughran2011liability}. 
The advantage of a numerical score is that it weighs words based on the sentiment content they convey. 
In addition, the sign of the sentiment deriving from the dependent words is adjusted to reflect the tone of the term of interest. 
For instance, we reverse the sign of the sentiment for words that are dependent on the term ``unemployment'' which has a negative connotation, or if we detect a negation in the chunk of text considered.
%% LOCATION
Another characteristic of this approach is that it aims at extracting from the text %from the sentence and the article 
the reference to a \textit{geographic location}. 
The goal is to include in the calculation of the sentiment only the text that references to the country we are targeting to monitor. 
In case no country can be identified in the text, we assume that it refers to the place where the journal is published. 
Overall, the FiGAS approach promises to deliver an indicator targeted to the topic of interest and that accurately measures sentiment due to its fine-grained and aspect-based nature. 
%We refer to \citep{consoli2021finegrained} for a detailed presentation of the algorithm.

% We then compute sentiment only on those terms and obtain a score between -1 and 1, as opposed to the typical approach of classifying words in the positive and negative categories \citep{loughran2011liability}. 
% The advantage of a numerical score is that we do not consider all positive and negative words equally, but weigh them based on the sentiment content they convey. In addition, the sign of the sentiment deriving from the dependent words is adjusted to reflect the tone of the term of interest. For instance, we reverse the sign of the tone for words that are dependent on the term ``unemployment'' which has a negative connotation. In addition, we also reverse the sentiment score if we detect a negation in the chunk of text considered.

%% ADD EXAMPLE SENTENCE
As an illustration, consider the sentence: ``\textit{The French economy has been experiencing its worst recession since 1968, while Italy entered into recession with a GDP drop ...}”, which appeared in the French newspaper \textit{La Tribune} on September $9^{th}$, 2020. 
If we are interested in extracting the sentiment about the topic \textit{Economy} with location \textit{France}, the FiGAS algorithm detects that only the first part of the sentence relates to this combination of topic and location. 
The algorithm also recognizes a specific semantic rule which, in this case, is that the topic of interest is followed by a direct object, and identifies the terms that characterize it, namely ``experience", ``worst" and ``recession". We assign a sentiment score to each of these words based on the proposed fine-grained dictionary and obtain an overall sentiment score of -0.98, which captures the negative outlook expressed in the sentence about the French economy.
\cite{consoli2021finegrained} provides a detailed presentation of the algorithm and a comparison with other popular sentiment analysis methods\footnote{The Python code to perform the FiGAS analysis is available at \url{https://github.com/sergioconsoli/SentiBigNomics} and a {\tt R} package is available at \url{https://github.com/lucabarbaglia/FiGASR}, where we also provide access to the proposed sentiment indicators.}.

In our application, we are interested in tracking the sentiment conveyed by news regarding the overall state of economic activity. 
We thus choose keywords that relate to the {\it Economy}, {\it Financial Sector}, {\it Inflation}, {\it Manufacturing}, and {\it Monetary Policy}. 
More specifically, we use the same terms as in \cite{barbaglia2020forecasting}, except for the {\it Monetary Policy} indicator where we include terms such as {\it European Central Bank} (ECB) and {\it Bank of England} (BOE), among others, to adapt the searches to the European scenario. The terms are: 
\begin{itemize}
 \item {\it Economy}: economy;
 \item {\it Financial Sector} ({\it finsector}): bank, derivative, lending, borrowing and combinations of [banking, financial] with [sector, commercial, and investment];
 \item {\it Inflation}: inflation; 
 \item {\it Manufacturing} ({\it manuf}): manufacturing and combinations of [industrial, manufacturing, construction, factory, auto] with [sector, production, output, and activity]; 
 %\item {\it Monetary Policy} ({\it monpol}): central bank, federal reserve, money supply, monetary policy, federal funds, base rate, and interest rate; 
 \item {\it Monetary Policy} ({\it monpol}): ecb, european central bank, boe, bank of england, money supply, monetary policy, base rate, interest rate, refinancing rate, marginal lending facility and deposit facility; 
 \item {\it Unemployment}: unemployment.
\end{itemize}

\textcolor{black}{
In the Online Appendix, we discuss in detail the main features of the algorithm and the fine-tuning required to adapt the analysis to the European news.
While the sentiment algorithm is the same used by \cite{barbaglia2020forecasting} to study the US case, its implementation differs in few aspects when working with non-English news from European newspapers.
Other than the keyword selection as detailed above for the case of \textit{monetary policy}, also the filtering on the most relevant geographic location based on named-entity recognition needs to be adapted.
Given that the sentiment algorithm is based on linguistic rules designed for the English language, we decided to translate the news article from Spanish, French, Italian, and German to English. 
We refer to the Online Appendix for further details on the robustness of the translation approach in sentiment analysis.}

\subsection{Forecasting models}

Our goal is to forecast annualized real GDP growth at the quarterly frequency at horizons that range from 15 days before the official release to approximately a year. The baseline specification that we consider is 
\begin{equation}
\label{eqn:eq.dl}
Y_t^d = \alpha_{h} + \eta_h S_{d-h} + X_{d-h}^{'} \beta_{h} + \epsilon^{d-h}_{t}~,
\end{equation}
where:
\begin{itemize}
 \item $Y_t^d$ represents the GDP growth for quarter $t$ that is released on day $d$; we use two indexes to track both the calendar time of the variable as well as the irregular release dates; 
 \item $S_{d-h}$ denotes a confidence or sentiment indicator available at the time the forecast is made on day $d-h$, that is, $h$ days before the official release date, $d$; the goal here is to accurately track the real-time flow of information that is available to forecasters;
 %\footnote{We consider the first release of the GDP variable that occurs, typically, 45 days after the end of the quarter.};
 \item $X_{d-h}$ indicates a vector of variables that are available on day $d-h$; in this vector we include lags of the GDP growth rate, current and lagged values of macroeconomic variables, and all survey and sentiment indicators other than $S_{d-h}$. In the empirical exercise, we track the release dates of the macroeconomic variables, allowing us to include in the vector $X_{d-h}$ only the values of the variables that are known to a forecaster on day $d-h$ when the forecast is produced;
 \item $\alpha_h$, $\eta_h$, and $\beta_h$ are the parameters to be estimated; we include the $h$ index to stress the fact that these coefficients are horizon-specific;
 \item $\epsilon_{t}^{d-h}$ is the forecast error for the quarter $t$ release of GDP growth that occurs on day $d$ based on the information available on day $d-h$.
\end{itemize}

The application involves data at mixed frequencies 
since the dependent variable (GDP growth) is available quarterly, while the predictors are at the monthly frequency. 
We handle the mixed frequency of the variables following the unrestricted mixed-data sampling (U-MIDAS) approach proposed by \citet{foroni2015unrestricted}. 
This approach consists of including the monthly variables and their lags as regressors in the forecasting regression. The method is unrestricted relative to the proposal by \citet{ghysels2007midas} and \citet{andreou2010regression} where the coefficients are assumed to follow a polynomial form. The dimension of the vector $X_{d-h}$ can be large\footnote{In our application, we include 9 monthly variables, more specifically 3 macroeconomic variables and 6 survey indicators. For each of the 9 variables we include at least 3 lags which, added to lags of the dependent variable, makes the number of parameters to estimate quite large.}, relative to the sample size. We use the U-MIDAS approach since it can be simply combined with a selection procedure that handles a situation in which the number of parameters is large relative to the sample size, such as lasso.

The parameter of interest in our analysis is $\eta_h$ that measures the effect on GDP growth of a change in the confidence or sentiment indicator. To conduct accurate inference on the parameter of interest $\eta_h$, we estimate Equation \eqref{eqn:eq.dl} using the double lasso methodology proposed by \citet{belloni2014inference}. 
The approach consists of a post-lasso estimation method in which the forecasting model is estimated using only the variables selected in a preliminary lasso selection step. 
However, the standard post-lasso procedure might lead to inconsistent estimates due to the possibility of eliminating relevant variables in the selection stage. \citet{belloni2014inference} propose to make the procedure more robust by running an additional lasso selection step in which the soft indicator $S_{d-h}$ is regressed on $X_{d-h}$. 
The union of the variables selected in these two preliminary lasso regressions is then used in the post-lasso step which delivers consistent and asymptotically normal estimates of $\eta_h$. In addition, in our in-sample analysis we test the significance of $\eta_h$ across multiple horizons $h$ which might lead to inaccurate inference. We thus correct the p-values for the multiple testing problem using the approach proposed by \citet{benjamini2006adaptive}.

\section{Data\label{sec_data}}
%% NEWS
Our data set consists of approximately 27 million articles published between January 1$^{st}$ 1995 and September 30$^{th}$ 2020 for the major economic and general purpose newspapers in Germany, France, Italy, Spain and the United Kingdom\footnote{Our sample consists of the following outlets: 
Süddeutsche Zeitung, Der Spiegel, Die Tageszeitung, Die Welt for Germany;
Le Monde, Le Figaro, Les Echos, La Tribune for France;
La Stampa, Corriere della Sera, Il Sole 24 Ore, Il Giornale, La Repubblica for Italy;
El Mundo, ABC, Expansión, La Vanguardia, Cinco Días, El País - Nacional for Spain;
The Times, The Guardian, Daily Mail, The Economist, Evening Standard, The Sunday Times, Observer for the UK.}. We obtain the articles from the Dow Jones Data, News and Analytics (DNA) commercial platform\footnote{The Dow Jones DNA platform is accessible at \url{https://professional.dowjones.com/developer-platform/}.}. 
The information provided with each article consists of title, snippet, body, date of publication, author(s), and topic categories. %classification. 
We collect articles for all DNA categories except for sport-related news.
%% TRANSLATE
Since the semantic rules in our FiGAS algorithm are designed for the English language, we translate all articles from German, French, Italian and Spanish to English using the neural machine translation service provided by the European Commission (EC)\footnote{More details about the \textit{eTranslation} service from the European Commission are available at \url{https://ec.europa.eu/info/resources-partners/machine-translation-public-administrations-etranslation_en}. We refer to the Appendix for an exploratory evaluation of the translation quality.}.

%% SURVEYS
In addition to the economic sentiment produced by the FiGAS algorithm, we consider survey expectations as an alternative source of real-time information about the state of the economy. We collect data for the BCS provided by the EC regarding six composite indicators, namely the Construction Confidence Index (CCI), the Industry Confidence Index (ICI), the Retail Sales Confidence Index (RCI), the Consumers Confidence Index (CSMCI), the Services Confidence Index (SCI), and the Economic Sentiment Index (ESI). The Survey starts in 1985 and the monthly indicators are released after the 20$^{th}$ day of each month. 

%% GDP and MACRO vars
Finally, we obtain macroeconomic variables from the ALFRED database of the Federal Reserve Bank of Saint Louis\footnote{The Saint Louis Fed ALFRED repository is accessible at \url{https://alfred.stlouisfed.org/}.}. In our analysis, we forecast the real GDP growth rate for the 5 European countries mentioned earlier and consider as predictors the industrial production index, the consumer price index and the unemployment rate at monthly frequencies.
We transform the first two indexes to percentage growth rates, while we take the first difference of the unemployment rate.
The database provides historical vintages and release dates starting in January 2011 for the monthly variables and, as of January 2016, for the GDP. In the empirical analysis, we forecast the flash estimate that is typically released 45 days after the end of the quarter \citep{eurostat2016euro}.
%\footnote{\textcolor{black}{In recent years the European statistical agencies have coordinated in providing also a preliminary flash estimate of GDP growth approximately 30 days after the end of the quarter, while the first estimate is only available after around 65 days .}}. 
We consider forecast horizons starting from 15 days before the official release up to approximately a year at intervals of 15 days for both the in-sample and out-of-sample exercise.

\subsection{Descriptive statistics}

\begin{figure}[]
 \centering
 \includegraphics[scale = 0.47]{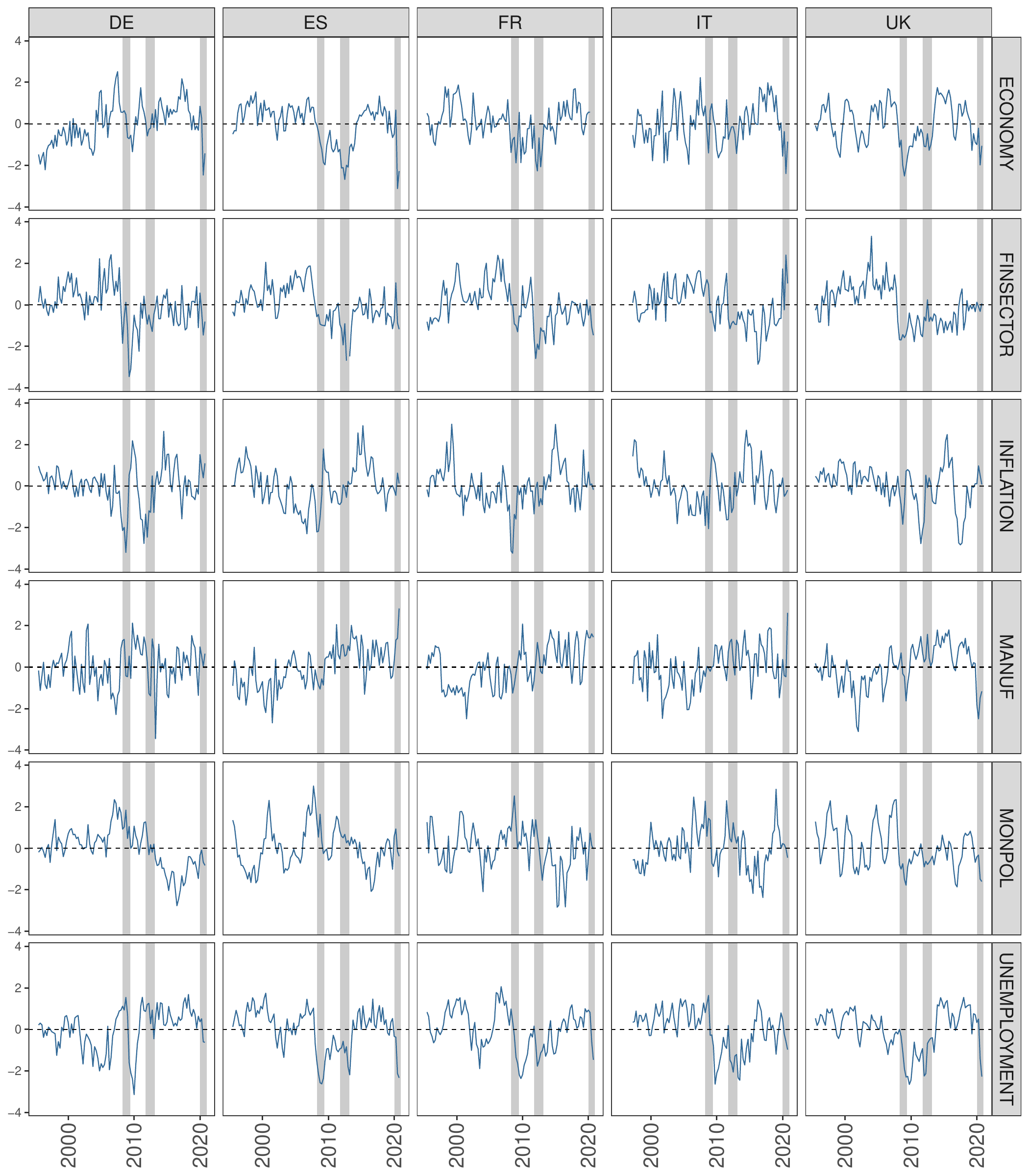} 
 \caption{\small Time series of the standardized news-based sentiment indicators for Germany (DE), Spain (ES), France (FR), Italy (IT), and the United Kingdom (UK). The sentiment is averaged within the quarter and sampled at the quarterly frequency.
 \textcolor{black}{The shaded areas represent the recessions established by the CEPR-EABCN business cycle dating committee.}
 }
 \label{fig:sent_ts}
\end{figure}

%% STATIONARITY OF THE SERIES (maybe check for multiple testing)
%% Plot the sentiment series + desnsities + correlations
Figure \ref{fig:sent_ts} shows the sentiment indicators of economic activity together with the recession periods as defined by the Centre for Economic Policy Research (CEPR) and Euro Area Business Cycle Network (EABCN) dating committee\footnote{\url{https://eabcn.org/dc/news}}. We aggregate the sentiment indicators from the daily to the monthly frequency by averaging the values within the month, and standardize the measures to have mean zero and variance one to make them comparable across countries and measures.
The {\it Economy} and {\it Unemployment} sentiments represent, for most countries, the more pro-cyclical indicators turning from positive to negative approximately at the beginning and at the end of the recessions.
%The sentiment about the {\it Economy} in Germany shows a different dynamics relative to the other countries since it is mostly positive during our sample period, although it declines during the contractions of the European economy. 
We also notice that the decline in sentiment during the 2008-09 economic slowdown is followed by a slow recovery and a further deterioration coinciding with the recession of 2011-2013. Germany seems to diverge from this pattern as there was a significant slump in sentiment about {\it Unemployment} during the first recessionary period, but only a minor decline in the {\it Economy} measure. 
In addition, the sentiment in German news about {\it Unemployment} is mostly negative in the early part of the sample and turns positive only in 2007, when the unemployment rate declined below 9\% for the first time since 1993. 
The sentiment about the {\it Financial Sector} is predominantly positive in the early part of the sample for all countries, but sharply declines during the 2008-2009 recession in Germany and the UK, and during the second recession in Spain, France, and Italy. 
The different behavior of the measure can be explained by news about the large exposure of German and British banks to the financial crisis in the United States, as opposed to banks in Spain, France, and Italy that were affected by exposure to the Greek debt crisis starting in 2011. 
Sentiment about {\it Manufacturing} is mostly negative during the 1990s and early 2000s in Spain, France, and the UK, while it is generally positive after 2013. Interestingly, we do not observe large swings in sentiment during the double-dip recession, probably due to the fact that negative news were mostly concentrated on the financial sector and the labor market.
%The {\it Inflation} indicator seems to vary on a wider range for the UK relative to the other European countries, and tends to become negative during the recessionary periods.
 The {\it Inflation} measure appears to be counter-cyclical, becoming negative during the expansion years of the early to mid-2000s, while being positive after 2013 when inflation was moderated by a recovering economy. 
%Similarly, the indicator for {\it Monetary Policy} appears to co-move more strongly among the continental countries while having different dynamics in the UK. This result is consistent with the fact that the four euro area countries are part of a monetary union which necessarily produces synchronization of their monetary-related news.

The behavior of the sentiment about {\it Monetary Policy} reflects the decisions of the ECB and the BOE. In particular, the ECB policy rate peaked between October 2000 and April 2001 to 3.75\%, followed by a steady decline toward 1\% that lasted until December 2005. After that, the ECB increased rates at steps of 0.25\% until reaching another peak of 3.25\% between July and October 2008. The overall level of sentiment in Spain, France, and Italy shows an increasing pattern which appears to follow rate increases, while declining as rates lowered. In addition, the peak in sentiment that is observed between the 2008-09 and 2011-13 recessions in these countries can also be related to the ECB policy decision to increase rates starting in April 2011, a decision that was ultimately reversed in November 2011. The {\it Monetary Policy} sentiment for the UK seems to have a break in mean and volatility in 2008 when its level and variation declined significantly. A possible explanation for this change of behavior is the shift in monetary policy by the BOE at the onset of the Great Recession. 
In the years between 1995 and 2008 the policy rate fluctuated between 3.5\% and 7.5\% and the sentiment cycles follow quite closely the dynamics of the rate\footnote{The fact that the sentiment for {\it Monetary Policy} co-varies positively with the level of the interest rate is due to the fact that the keyword {\it interest rate} has a positive tone which, associated with words like {\it increase} or {\it raise}, provides an overall positive sentiment. More specifically, the word {\it rate} has neutral sentiment while {\it interest} is positive in the dictionary. Our algorithm propagates the sentiment of {\it interest} to {\it rate} so that the combination {\it interest rate} has a positive sentiment. Similarly, the words {\it increase} and {\it raise} carry positive sentiment and when they refer to {\it interest rate} reinforce its positive sentiment. The positivity of {\it interest rate} is debatable, in particular in a macroeconomic and monetary policy context. However, an {\it increase} in {\it interest rate} represents a price that benefits lenders and depositors while hurting borrowers. This example demonstrate the complexity of assigning sentiment to words and sentences due to the aspect-specific meaning of many words. On the other hand, since we are using these indicators in a regression context, the negative macroeconomic effect of an increase in rates (associated with positive sentiment) can be accommodated by the negative coefficients.}. The BOE response to the Great Recession was to lower the rate between September 2008 and March 2009 from 5\% to 0.5\%, where it stayed for several years. The {\it Monetary Policy} sentiment experienced a decline as well and much lower variability, probably due to the fact that the rate was fixed at 0.5\% in that period.

%\begin{center} {\bf Figure~\ref{fig:sent_ts} approximately here}\end{center}

% \begin{figure}[ht]
% \centering
% \includegraphics[scale = 0.47]{Figures/DNA_Sentiment_Europe_sent=av_quarter.pdf} %% UPDATE with MonPol_EU
% \caption{\small Time series of the six news-based sentiment indicators for Germany (DE), Spain (ES), France (FR), Italy (IT), and the United Kindgom (UK). The sentiment is averaged within the quarter and sampled at the quarterly frequency.}
% \label{fig:sent_ts}
% \end{figure}

%\begin{figure}[ht]
% \centering
% \includegraphics[scale = 0.5]{Figures/DNA_Sentiment_Europe_gdp&sent_quarter.pdf}
% \caption{\small Time series of the six sentiment indicators for Germany (DE), Spain (ES), France (FR), Italy (IT), and the United Kindgom (UK). The sentiment represents a 30-day moving average of the daily series and it is sub-sampled at the monthly frequency. }
% \label{fig:sent_ts}
%\end{figure}

\begin{figure}[]
 \centering
 \includegraphics[scale = 0.47]{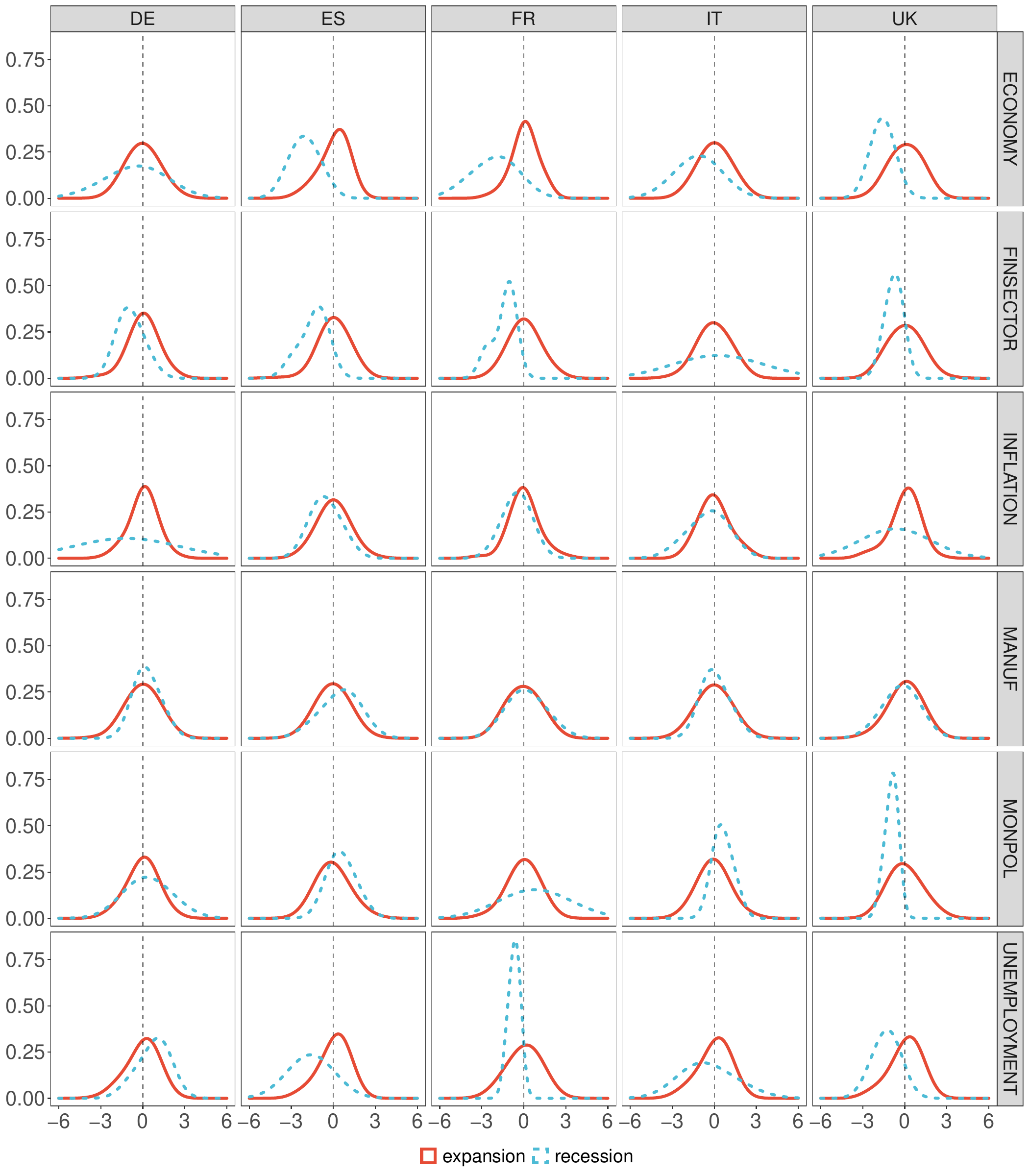} 
 \caption{\small Kernel density estimates of the standardized sentiment indicators during expansions and recessions by country and topic.
 \textcolor{black}{Expansions and recessions are based on the classification of the CEPR-EABCN business cycle dating committee.}}
 \label{fig:sent_dens}
\end{figure}

%% DENSITIES
The dependence of the scaled sentiment measures on the business cycle is also apparent in Figure \ref{fig:sent_dens} which shows the kernel density estimate of the indicators during expansionary and recessionary periods.
%\footnote{{\color{black} Country-specific expansionary and recessionary periods are obtained from the OECD Composite Leading Indicators repository available at \url{https://www.oecd.org/sdd/leading-indicators/oecdcompositeleadingindicatorsreferenceturningpointsandcomponentseries.htm}.}}. 
In addition to the pro-cyclical nature of the \textit{Economy} and \textit{Unemployment} sentiments that is also %apparent 
evident in these graphs, another interesting insight %fact 
is the larger variability of the {\it Inflation} indicator during recessions, relative to expansions in Germany and the UK. 
A possible explanation is that, during recessions, there is more uncertainty about the impact on inflation of stimulative fiscal and monetary policies and, in general, on the state of the economy. 
%As discussed earlier, the lower level of {\it Monetary Policy} sentiment in the UK follows the sudden rate drop below 1\% following the financial crisis. 
Another indicator that experienced a significant fall during recessions in most countries was the {\it Financial Sector}. This is not surprising as a result of the double-dip recession developed as a consequence of the significant deterioration of financial conditions in many European economies.
% \begin{figure}[ht]
% \centering
% \includegraphics[scale = 0.47]{Figures/DNA_Sentiment_Europe_density_sent=av.pdf} 
% %\includegraphics[scale = 0.47]{Figures/DNA_Sentiment_Europe_density_sent=av_OECDbusinesscycle_kernel.pdf} 
% \caption{\small Kernel density estimates of the sentiment indicators during expansions and recessions by country and topic. 
% {\color{black} Country-specific expansionary and recessionary periods are obtained from OECD Composite Leading Indicators.}}
% \label{fig:sent_dens}
% \end{figure}

%\begin{center} {\bf Figure~\ref{fig:sent_dens} approximately here}\end{center}

% \begin{figure}[ht]
% \centering
% %\includegraphics[scale = 0.47]{Figures/DNA_Sentiment_Europe_density_sent=av.pdf} 
% \includegraphics[scale = 0.47]{Figures/DNA_Sentiment_Europe_density_sent=av_OECDbusinesscycle_kernel.pdf} 
% \caption{\small Kernel density estimates of the sentiment indicators during expansions and recessions by country and topic. 
% {\color{black} Country-specific expansionary and recessionary periods are obtained from OECD Composite Leading Indicators.}}
% \label{fig:sent_dens}
% \end{figure}

\begin{figure}[]
 \centering
 %\includegraphics[scale=0.5]{Figures/DNA_Sentiment_Europe_correlation_sent=av.pdf}
 %\caption{\small Correlation coefficients of the sentiment indicators across countries. }
 %\includegraphics[scale=0.57]{Figures/DNA_Sentiment_Europe_correlation_sent=av_GGALLY.pdf} %% Updated with MonPol_EU
 \includegraphics[scale=0.57]{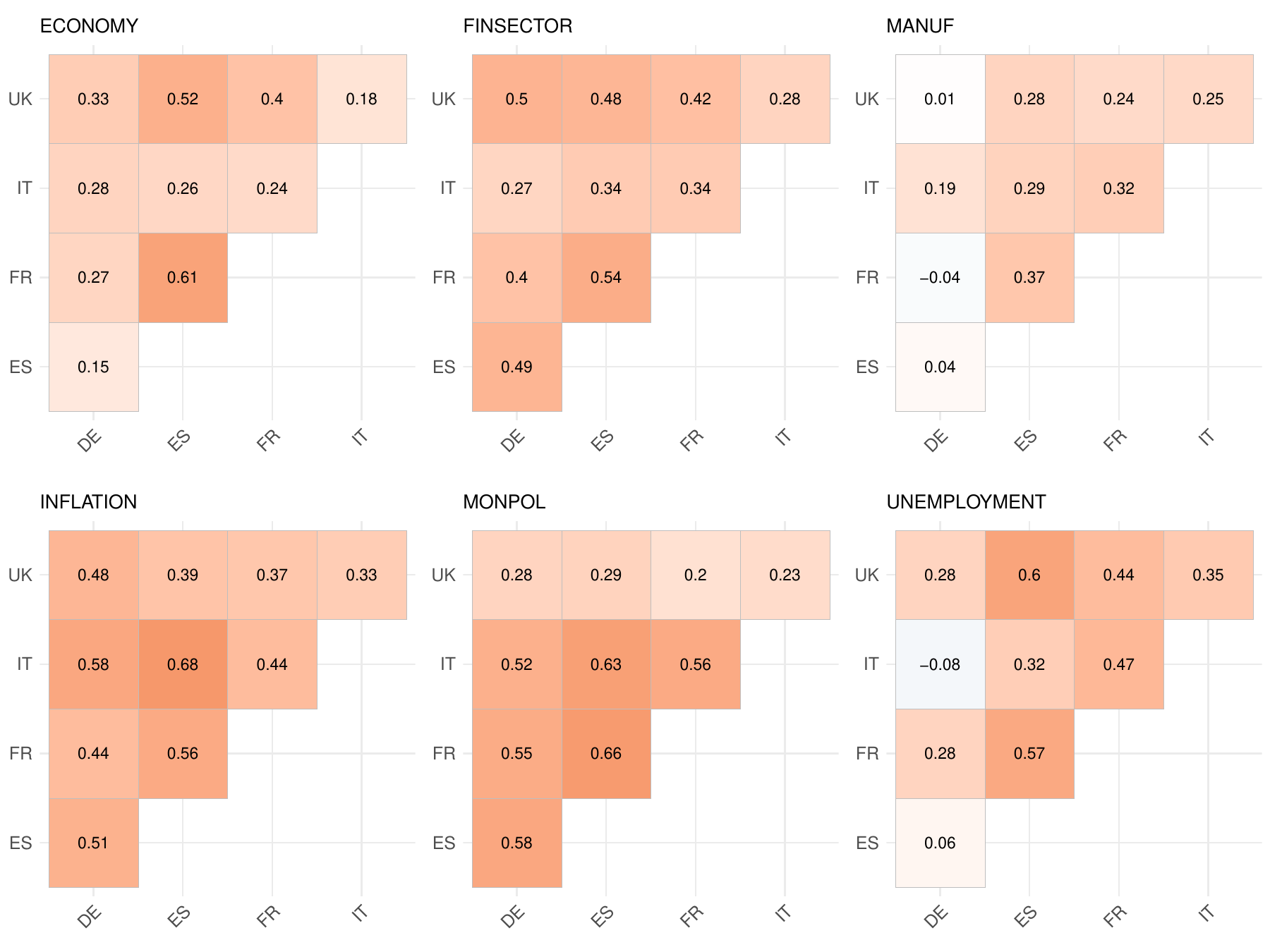} %% Updated with MonPol_EU
 \caption{\small Correlation coefficients of the sentiment indicators across countries. 
 \textcolor{black}{Darker colors indicate larger correlation in absolute value.}
 }
 \label{fig:sent_corr}
\end{figure}

%% CORRELATION
The time series behavior of the indicators in Figure \ref{fig:sent_ts} seems to suggest a significant degree of co-movement in sentiment across European countries. 
To evaluate this commonality, Figure \ref{fig:sent_corr} provides the estimates of the correlation between the sentiment indicators across countries calculated at the quarterly frequency. 
The correlation across indicators and countries ranges between -0.08 ({\it Unemployment} in Germany and Italy) and 0.68 ({\it Inflation} in Italy and Spain). 
Correlation in the sentiment for {\it Manufacturing} is typically lower among all pairs of countries, as opposed to correlation regarding {\it Financial Sector}, where sentiment is more correlated. 
Interestingly, there are large co-movements in sentiment among the euro area countries about {\it Monetary Policy} and \textit{Inflation} driven by the common flow of news regarding monetary events. 
Overall, Germany and the United Kingdom seem to be less correlated on the {\it Economy} and {\it Manufacturing} indicators while France, Italy and Spain constitute a more correlated block. 
{\color{black} In the Online Appendix we also investigate the Granger-causality patterns between survey's confidence and news sentiments.  The evidence suggests the existence of bidirectional relations, with news having significant additional explanatory power with respect to traditional indicators. }

%\begin{center} {\bf Figure~\ref{fig:sent_corr} approximately here} \end{center}

%\end{comment}

\section{In-sample analysis\label{sec_insample}}

\begin{figure}[]
 \centering
 \includegraphics[scale = 0.46]{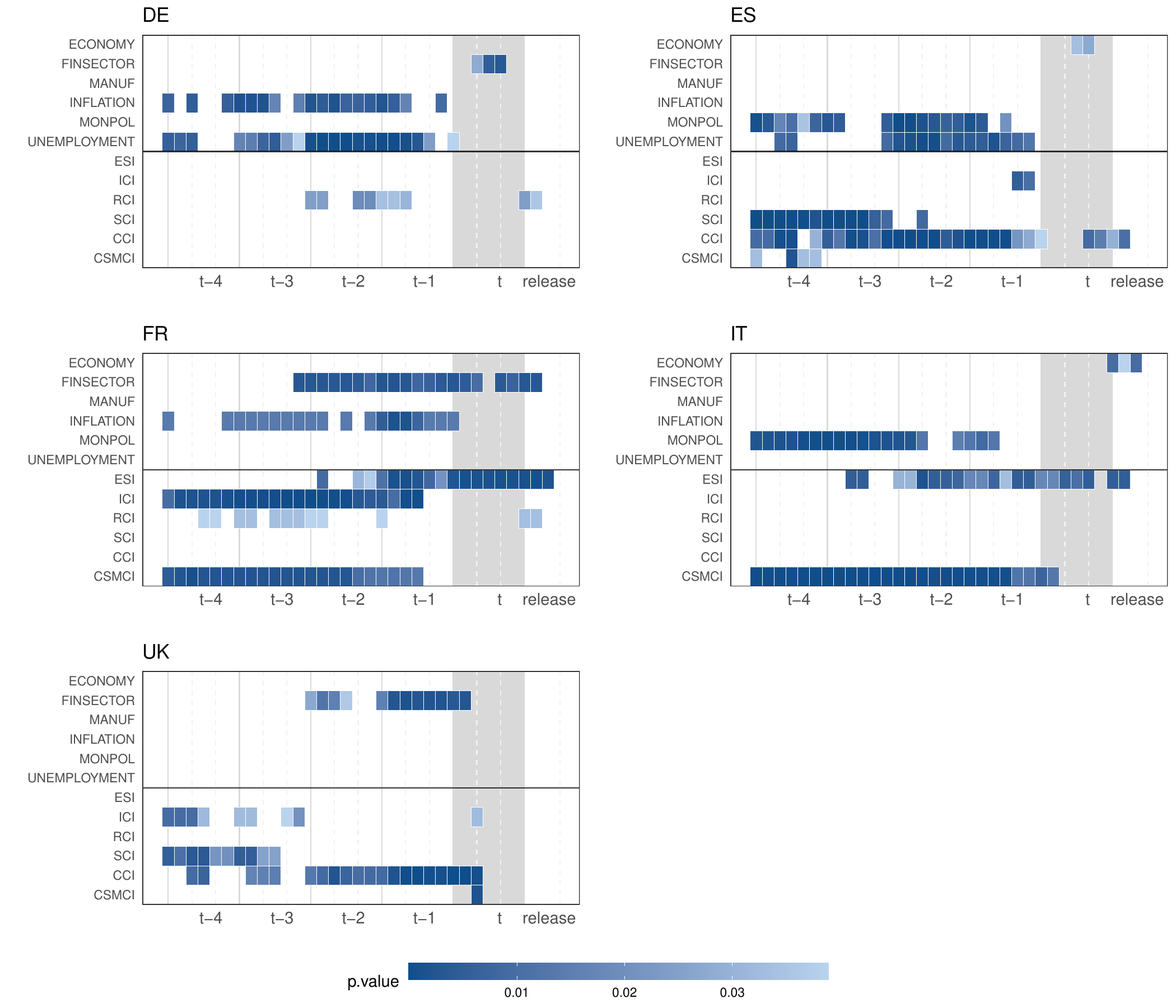} 
 \caption{ Statistical significance of the survey and sentiment measures as predictors of GDP growth by country with p-values corrected for multiple testing across horizons. 
\textcolor{black}{ The grey area represents the quarter being forecast and {\it release} indicates the release date.} 
 %The shadowed area represents the quarter being nowcast. 
 The $x$-axis reports the horizon $h$, which ranges from 15 days before the release date to approximately 4 quarters ahead at intervals of 15 days. The color of the tile represents the p-value of the coefficient of the survey or sentiment indicators $\eta_h$ in Equation \eqref{eqn:eq.dl}: the darker the tile, the smaller the p-value.}
 \label{fig:sent_insample}
\end{figure}

Figure \ref{fig:sent_insample} shows the statistical significance of the survey and sentiment indicators as predictors of GDP growth at the different forecasting horizons.
The figure shows the p-value adjusted for multiple testing for the $\eta_h$ coefficient in Equation \eqref{eqn:eq.dl}, obtained using the double lasso approach by \citet{belloni2014inference}. 
%% SURVEYS
Regarding the EC survey indicators, we find that the consumer confidence measured by the CSMCI is a strong predictor in France and Italy, and mostly at long forecasting horizons. 
Other survey expectations that are significant at medium and long horizons are for the industrial sector (ICI for France), the construction sector (CCI, for Spain and the UK), and services (SCI for Spain and the UK). 
The broader ESI index is significant at 5\% for France and Italy at horizons of one quarter or shorter. 
%Instead, the survey indicators that are significant at nowcasting horizons are CCI (Spain), RCI (Germany, and Italy), and ESI (France and Italy). 

%% NEWS
Regarding the news-based sentiment measures, we find for most countries a clearly defined pattern in which only a few sentiment measures provide incremental predictive power relative to the macroeconomic variables and the confidence indicators. 
In particular, we notice that {\it Unemployment} is strongly significant to forecast GDP at the one-quarter ahead and longer horizons in Germany and Spain. 
The tone embedded in the news regarding {\it Monetary Policy} seems particularly relevant in Spain and Italy at similar forecast horizons, while the {\it Financial Sector} indicator seems especially useful to forecast GDP in France and the UK, although for Germany it is most relevant at nowcasting horizons.
Interestingly, the \textit{Manufacturing} sentiment is not selected as a predictor for GDP growth in any of the countries.
Finally, the sentiment about the {\it Economy} appears to provide predictive power at short horizons, such as in the case of Spain and Italy.

%\begin{center} {\bf Figure~\ref{fig:sent_insample} approximately here}\end{center}

% \begin{figure}[ht]
% \centering
% % \includegraphics[scale = 0.48]{Figures/figasEU_insample_doublelasso_significance_MT_pval05.pdf}
% \includegraphics[scale = 0.47]{Figures/figasEU_insample_doublelasso_significance_MT_pval10_asymmetric_SM.pdf}
% \caption{ Statistical significance of the survey and sentiment measures as predictors of GDP growth by country \textcolor{black}{after the correction for multiple testing across horizons}. The shadowed area represents the quarter being nowcasted. 
% The $x$-axis reports the horizon $h$, which ranges from 15 days before the release date to approximately 4 quarters ahead at intervals of 15 days. The color of the tile represents the p-value of the coefficient of the survey or sentiment indicators $\eta_h$ in Equation \eqref{eqn:eq.dl}: the darker the tile, the smaller the p-value.}
% \label{fig:sent_insample}
% \end{figure}

Overall, the evidence suggests that both surveys and news-based sentiment indicators are useful to predict GDP growth. 
While the survey measures are more focused on capturing the expectations of consumers and businesses, the sentiment indicators aggregate the more general tone of the economic discussion regarding the state of the economy. 
An advantage of using news is that they allow estimating sentiment on a broader set of topics of interest.
The inclusion of news-based sentiment measures about the {\it Financial Sector} and {\it Monetary Policy}, which are not included in the EC survey, seems to be a relevant factor in predicting GDP growth in Spain, France and the UK. 
The analysis also shows that sentiment indicators that capture different aspects of economic activity might be useful to accommodate the local attitudes in each country. 
For example, the German public might be more sensitive to news that discuss the current and future outlook of inflation, which thus conveys sentiment that is more informative to forecast GDP. 
On the other hand, news articles reporting on the health of the banking and financial sectors might draw more attention given their relevance for the UK economy, and thus carrying higher predictive content for the GDP. 

% \begin{figure}[ht]
% \includegraphics[scale=0.47]{Figures/figasEU_insample_doublelasso_significance_conditioning.pdf}
% \caption{In-sample double lasso}
% \label{fig:sent_insample}
% \end{figure}

\section{Out-of-sample analysis\label{sec_oos}}

The aim of this section is to evaluate the out-of-sample performance of the sentiment indicators to forecast the real GDP growth of the five European countries we are considering. 
In particular, we would like to evaluate the incremental predictive content of the news-based sentiment measures relative to the information already provided by the macroeconomic indicators and the confidence indexes of the EC. 
We perform this by including in the vector $X_{d-h}$ in Equation \eqref{eqn:eq.dl} the lags of the dependent variable, and the current and lagged values of the macroeconomic variables and of the survey indicators. 
Our benchmark forecasting model, which we call ARX, consists of the post-lasso model that includes only the variables selected following the double lasso procedure discussed in Section \ref{sec_methodology}. 
Instead, the ARXS model augments the baseline specification with a news-based sentiment measure and produces forecasts that are evaluated against the ARX benchmark forecasts. 
Should we find higher predictive accuracy of the ARXS forecasts relative to the ARX forecasts, it would indicate that the sentiment measures carry genuine and relevant information that is incremental to that provided by the macroeconomic and survey indicators. 

%% setup
The forecast period ranges from the first quarter of 2007 to the fourth quarter of 2019, for a total of 13 years of quarterly forecasts. The choice of starting the out-of-sample period in the first quarter of 2007 is motivated by the intention to include in the evaluation the double-dip recession that occurred in Europe between 2008 and 2013.
We use the same forecast horizons $h$ employed for the in-sample analysis that ranges from 15 to 495 days before the release date, at intervals of 15 days. The forecasting models are estimated using an expanding window that starts in the first quarter of 1997. The size of the samples for both estimation and forecast evaluation are quite short, and this might affect negatively the accuracy of our forecasts. Nevertheless, we believe that it is still useful and informative to perform an out-of-sample exercise to measure %have a sense of 
the relative performance of the competing forecasting models. Besides, we can always rely on the in-sample results presented in Section~\ref{sec_insample} if we consider that the out-of-sample period is structurally different from the rest of the sample. In addition to the ARX and ARXS forecasts, we consider the case of a pooled forecast obtained by averaging the six ARXS forecasts, which we refer to as the \textit{Average} forecast.

\begin{figure}
 \centering
 \includegraphics[scale=0.47]{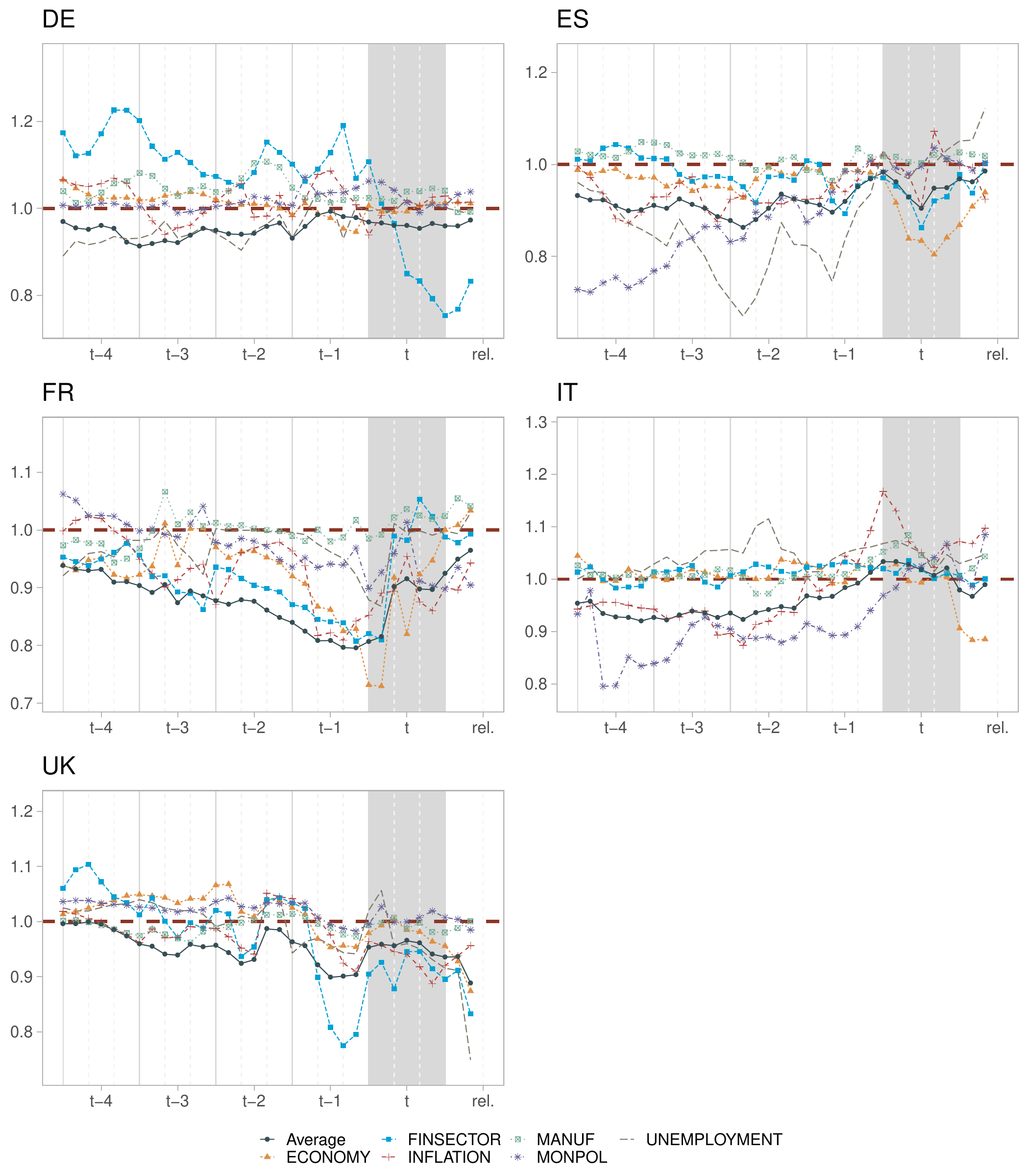}
 \caption{ Ratio of the MSFE for the ARXS specification relative to the ARX benchmark across horizons. The grey area represents the quarter being forecast and {\it rel.} indicates the release date.  The $x$-axis reports the horizon $h$, which ranges from 15 days before the release date to approximately 4 quarters ahead at intervals of 15 days.
 }
 \label{fig:MSFE_REL}
\end{figure}

%% DISCUSSION OF RELATIVE MSPE
First, we evaluate the performance of the sentiment indicators in terms of reduction of the Mean Square Forecast Error (MSFE) of the ARXS forecasts relative to the ARX forecasts at each horizon $h$. Figure~\ref{fig:MSFE_REL} shows that in several countries there is a considerable reduction in MSFE that exceeds 20\% at some horizons. Similarly to the in-sample results, we find that the sentiment measures play a larger role in increasing the accuracy of the forecasts at long horizons, rather than in nowcasting. However, the patterns are quite different across countries. 
For Spain we find that the majority of the sentiment measures contribute to increase forecast accuracy, although it tends to deteriorate approaching the release date, with the exception of the {\it Economy} indicator. The largest reductions in MSFE are obtained when considering the {\it Unemployment}, {\it Financial Sector} and {\it Monetary Policy} indicators in quarters $t-2$ and $t-3$. France has a similar behavior, although the sentiments achieve their maximum gains during quarter $t-1$, while deteriorating quickly in nowcasting. In the case of Italy, the results indicate that the {\it Monetary Policy} provides significant gains in predictability between quarter $t-4$ and $t-2$, while {\it Inflation} and the {\it Average} provide more modest improvements. In both the UK and Germany the only measure that provides a gain larger than 10\% is the {\it Financial Sector} at some selected horizons. In terms of nowcasting, the most significant result is the {\it Financial Sector} indicator in Germany, the {\it Economy} in Spain and Italy, and several measures in France and the UK, although with more limited gains. 
\textcolor{black}{
Overall, while high-frequency macroeconomic indicators are often found to  be powerful predictors at nowcasting horizons  \citep{mccraken2021realtime},  our results provide evidence that economic sentiments are relevant to improve forecast accuracy in particular at forecast horizons of 1 to 4 quarters ahead. }

%\begin{center} {\bf Figure~\ref{fig:MSFE_REL} approximately here}\end{center}

%% MultiHorizon SPA
Figure~\ref{fig:MSFE_REL} shows that there are instances of significant improvements in forecast accuracy when considering the sentiment measures as predictors of GDP. To evaluate the statistical significance of these improvements, we apply the multi-horizon average Superior Predictive Ability (aSPA) test proposed by \citet{quaedvlieg2021multi}. In order to evaluate separately the performance of the sentiment measures at the nowcasting and forecasting horizons, we perform the test on the set of horizons that are smaller or larger than 165 days, respectively\footnote{The choice of this threshold corresponds approximately to a quarter plus the typical 45 days of delay in the official release of the GDP data.}. The multi-horizon test provides a more robust performance assessment of the competing models, although it does not provide an insight on the horizons, if any, in which the sentiment-based forecasts outperform the macro/survey-based forecasts.

\begin{table}[]
\centering
\setlength{\tabcolsep}{16pt}
\begin{tabular}[t]{lccccc}
\hline\hline\\
Sentiment & DE & ES & FR & IT & UK\\
\hline \\
& \multicolumn{5}{c}{\it Nowcasting}\\ \\
ECONOMY & 0.111 & \bf{0.012} & \it{0.060} & 0.413 & \bf{0.036}\\
FINSECTOR & 0.105 & \bf{0.039} & 0.130 & 0.938 & \bf{0.030}\\
INFLATION & 0.643 & 0.321 & \bf{0.046} & 0.991 & \it{0.071}\\
MANUFACTURING & 0.973 & 0.987 & 0.769 & 0.975 & \bf{0.039}\\
MONETARY POLICY & 0.893 & \bf{0.021} & \it{0.080} & 0.598 & 0.320\\
UNEMPLOYMENT & 0.215 & 0.400 & \bf{0.037} & 0.962 & \it{0.066}\\ \\
AVERAGE & \bf{0.018} & \bf{0.009} & \bf{0.013} & 0.922 & \bf{0.032}\\
\\
& \multicolumn{5}{c}{\it Forecasting}\\ \\
ECONOMY & 0.965 & \bf{0.032} & \it{0.096} & 0.967 & 0.970\\
FINSECTOR & 0.889 & 0.323 & \bf{0.036} & 0.806 & 0.482\\
INFLATION & 0.631 & \bf{0.012} & 0.264 & \bf{0.020} & 0.403\\
MANUFACTURING & 0.985 & 0.937 & 0.258 & 0.841 & 0.316\\
MONETARY POLICY & 0.934 & \bf{0.016} & 0.321 & \bf{0.013} & 0.978\\
UNEMPLOYMENT & \it{0.081} & \bf{0.020} & \bf{0.019} & 0.923 & 0.664\\ \\
AVERAGE & \bf{0.025} & \bf{0.014} & \bf{0.028} & \bf{0.003} &\bf{0.032}\\
\\ \hline\hline
\end{tabular}
\caption{Multi-horizon test for equal predictive accuracy when considering exclusively the {\it nowcast} horizons (top part), and only the {\it forecast} horizons (bottom part). The benchmark model is the ARX and the alternative is the ARXS specification that augments the ARX with a news-based sentiment measure. The Average forecast is obtained from averaging the ARXS forecasts. The value in the table represents the p-value for the one-sided hypothesis that the forecasting models in the column outperform the ARX forecasts: p-values indicating significance at 5\% are denoted in bold and at 10\% in italics.
%p-values in bold indicate significance at least at 10\% level.
}
\label{tab:multi_forecast}
\end{table}

% %%%% NEW TAB
% \begin{table}[ht]
% \centering
% \setlength{\tabcolsep}{16pt}
% \begin{tabular}[t]{lccccc}
% \hline\hline\\
% Sentiment & DE & ES & FR & IT & UK\\
% \hline \\
% & \multicolumn{5}{c}{\it Nowcasting}\\ \\
% ECONOMY & 0.127 & \textbf{0.010} & \textbf{0.077} & 0.375 & \textbf{0.038}\\
% FINSECTOR & \textbf{0.095} & \textbf{0.040} & 0.113 & 0.949 & \textbf{0.029}\\
% INFLATION & 0.633 & 0.328 & \textbf{0.043} & 0.990 & \textbf{0.070}\\
% MANUFACTURING & 0.973 & 0.988 & 0.771 & 0.975 & \textbf{0.050}\\
% MONETARY POLICY & 0.980 & 0.142 & 0.127 & 0.424 & 0.415\\
% UNEMPLOYMENT & 0.204 & 0.374 & \textbf{0.030} & 0.961 & \textbf{0.063}\\
% \\
% AVERAGE & \textbf{0.017} & \textbf{0.010} & \textbf{0.031} & 0.916 & \textbf{0.040}\\
% \\
% & \multicolumn{5}{c}{\it Forecasting}\\ \\

% ECONOMY & 0.966 & \textbf{0.035} & 0.114 & 0.948 & 0.952\\
% FINSECTOR & 0.903 & 0.317 & \textbf{0.037} & 0.841 & 0.479\\
% INFLATION & 0.607 & \textbf{0.007} & 0.264 & \textbf{0.012} & 0.403\\
% MANUFACTURING & 0.989 & 0.939 & 0.263 & 0.851 & 0.287\\
% MONETARY POLICY & 0.917 & \textbf{0.021} & 0.411 & \textbf{0.020} & 0.986\\
% UNEMPLOYMENT & \textbf{0.086} & \textbf{0.021} & \textbf{0.026} & 0.923 & 0.674\\
% \\
% AVERAGE & \textbf{0.019} & \textbf{0.022} & \textbf{0.031} & \textbf{0.004} & \textbf{0.038}\\
% \\ \hline\hline
% \end{tabular}
% \caption{DELETE!!}
% \label{tab:multi_forecast}
% \end{table}

%% MultiHorizonSPA
The results for the multi-horizon test are provided in Table \ref{tab:multi_forecast}, with nowcast and forecast horizons in the top and bottom parts of the table, respectively. 
The values in the table represent the p-values for the one-sided hypothesis that the ARX macro/survey-based forecasts are outperformed by the ARXS sentiment-based forecasts.
The results indicate that the {\it Average} provides more accurate nowcasts for all countries except Italy at nowcasting horizons. 
%The result is confirmed also at the forecasting horizons where the sentiment-based forecasts are significantly more accurate than the macro/survey-based ones in all countries. 
The performance of the forecasts based on the individual sentiment indicators is mixed. The sentiment about the {\it Financial Sector} is significant in nowcasting GDP growth in Spain and the UK. We believe this finding might be driven by the rapid deterioration of financial conditions during the double-dip recession in Europe that is captured in real-time by the sentiment measures, and only reflected with a long delay in the macroeconomic and confidence indicators. 
Other significant indicators to nowcast GDP are the {\it Economy} sentiment for Spain and the UK, {\it Inflation} and {\it Unemployment}  in France, {\it Monetary Policy} in Spain, and {\it Manufacturing} in the UK. 
When considering the longer horizons, we find that the out-of-sample results are more consistent with the in-sample evidence discussed in the previous section. 
In particular, the sentiment about {\it Unemployment} is more accurate (relative to the benchmark) for Spain and France, while the {\it Monetary Policy} sentiment measure results to be significant for both Spain and Italy. 
The sentiment about {\it Inflation} provides predictive accuracy for Spain and Italy, whereas the {\it Financial Sector} sentiment shows to be significant to forecast French GDP relative to the benchmark forecasts. 
Interestingly, none of the sentiment measures is significant at 10\% to forecast British GDP. 
Similarly to the nowcasting case, we find that averaging the sentiment-based forecasts delivers more accurate forecasts relative to the macro/survey-based forecasts for all considered countries. 
Overall, the results support the in-sample evidence, showing that sentiment measures based on textual analysis of news articles provide a useful addition to the tools of economic forecasters both at short %as well as 
and long horizons. 

\begin{figure}[]
  \includegraphics[scale=0.45]{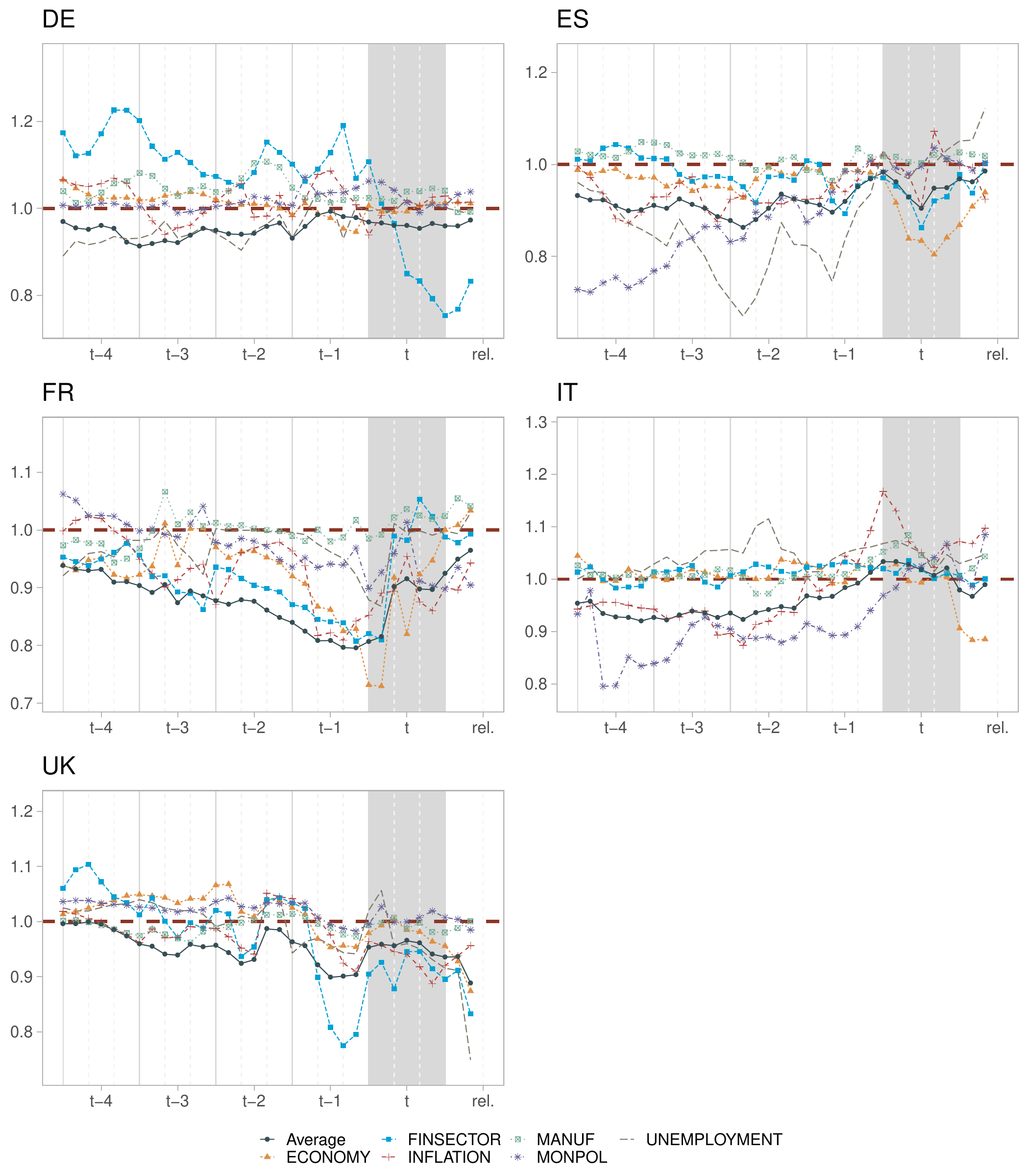}
 \caption{Fluctuation test for the ARXS and AVERAGE forecasts at four representative horizons. The size of the rolling window is 20\% of the out-of-sample period. The test statistic is centered at the midpoint of the rolling window and the grey areas represent the  CEPR-EABCN recession periods. {\color{black} The dashed line represents the one-sided critical value at 10\%.}} %% Updated with MonPolEU and corrected from duplicates
 \label{fig:fluct}
\end{figure}

{\color{black}
The evidence in \cite{barbaglia2020forecasting} for the case of the US indicates that the contribution of news-based sentiment measures 
to increase the forecast accuracy is episodic, that is, limited to specific periods of time during the sample. To assess this hypothesis in the context investigated here, we perform the fluctuation test proposed by \citet{giacomini2010forecast} which allows to evaluate the variation over time in the relative performance of the forecasts. Figure~\ref{fig:fluct} shows the fluctuation test at three horizons that correspond, approximately, to the end of the nowcasting quarter,  the previous quarter, and four quarters ahead the target. The lines represent the predictive accuracy of the ARXS forecasts, relative to the benchmark, using the six sentiment indicators that have been computed on a rolling window of 20\% of the out-of-sample period. We find that the earlier results that the {\it Economy} sentiment is useful to forecast in Germany is probably due to its high accuracy during the second recessionary period in 2011-2013. In addition, for Spain, Italy, and France the higher accuracy of the sentiment indicators is concentrated, to a large extent, during  the recessionary period. These results are, to some extent, consistent with the conditional predictive accuracy test discussed in the Online Appendix, even though the latter might have higher power to detect non-smooth breaks relative to the fluctuation test. For the UK, only at the shortest horizons the {\it Economy} measure provides significant accuracy relative to the benchmark around 2015. }

%% ADDITIONAL RESULTS in the APPENDIX

The Online Appendix reports additional analysis that attest the quality and  robustness of the paper's results. 
First, we test the significance of the forecast gains using the predictive accuracy test by {\color{black} \cite{giacomini2006tests}}, as well as, its conditional version by \cite{granziera2019predicting}. The proposed sentiment measures provide more accurate forecast during recessions and in mid-long horizons.
%\sout{Second, we confirm that the accuracy gains are, to a large extent, concentrated during  recessionary periods using the fluctuation test of \cite{giacomini2010forecast}}.
{\color{black} In addition}, we investigate whether the effect of the changes in sentiment on the macroeconomic outcome is non-linear. To this extent, we estimate an alternative model to Equation \eqref{eqn:eq.dl} where we split the sentiment measure $S$ in positive and non-positive. 
The out-of-sample performance of the non-linear model matches the one of the linear case reported in the main paper, hinting that non-linearity might  play a minor role.
\textcolor{black}{
Finally, we discuss the performance of the
proposed forecasting model during the Covid-19 pandemic: the results indicate that the forecasts based on the sentiment indicators are useful to reveal the rapid decline in the second quarter of 2020 and the rebound in the following quarter. }
%However, the forecasts are, in most countries, significantly smaller than the value that actually realized. 
%This result is, at least partly, to be expected, since the shock was of a magnitude that had never occurred in the sample period that we are considering

%% APPENDIX : other variables than GDP\\
%\sout{Finally, the Appendix reports the forecasting and nowcasting performance of the proposed sentiment indicators when focusing on macroeconomic variables other than GDP, namely the unemployment rate, the industrial production index and the consumer price index (CPI).}

\section{Forecasting other macroeconomic variables}
\label{sec:other_macrovariables}

\textcolor{black}{
In this section we discuss the robustness of our findings for GDP when forecasting other macroeconomic variables. The variables that we consider are the monthly change in the unemployment rate and the monthly growth rates of the Industrial Production Index (IPI) and the Consumer Price Index (CPI). The model setup is the same as earlier, although we now include  the GDP growth rate as a predictor. Tables~\ref{tab:multi_forecast_UNEMP} to \ref{tab:multi_forecast_CPI} provide the p-values of the multi-horizon forecast accuracy test proposed by \cite{quaedvlieg2021multi}. Since the variables being forecast are now at the monthly frequency, we define the horizons within the quarter of the release month as {\it short-run} and the ones in the following quarters as {\it long-run} horizons.}

\bigskip\noindent
%% UNEMP
\textcolor{black}{
Table \ref{tab:multi_forecast_UNEMP} reports the results for the change in the unemployment rate. 
The \textit{Average} forecast confirms its high accuracy also in this case at both short and long horizons for most countries.  The sentiment about {\it Unemployment} delivers more accurate short-run forecasts for Germany and for Spain. Other indicators that are also significant in the short-run are the {\it Economy} for France and the UK, while the {\it Manufacturing} and {\it Monetary Policy} measures matter only in the case of Spain. 
Instead, at longer horizons, 
sentiment about {\it Inflation} outperforms (at 10\%) the benchmark forecasts for Spain and France.
Similarly to the short horizon case, sentiments about \textit{Monetary Policy} and \textit{Unemployment} are significant in Spain.} 

\bigskip\noindent
{\color{black} Interestingly, Table \ref{tab:multi_forecast_INDPRO} shows that the pattern of significance when forecasting the IPI growth rate is similar to the case of the unemployment rate.
In particular, we find that at short horizons the sentiment about {\it Monetary Policy} and {\it Unemployment} contribute to produce more accurate forecasts in the case of Spain, as well as the  {\it Economy} indicator for the UK. 
However, at the longer horizons there are only a few indicators that are significant to predict the IPI growth rate, in particular {\it Monetary Policy}, while the average forecast is significant at all horizons for most countries.}

\bigskip\noindent
{\color{black} Finally, Table \ref{tab:multi_forecast_CPI} reports the results when forecasting the CPI inflation.
In this case, the sentiment measure about \textit{Inflation} is an important predictor at short horizons for Spain, Italy, and the UK, but not at longer horizons. Sentiment about {\it Unemployment} is significant at all horizons to forecast inflation in the UK, while the {\it Financial Sector}  is a useful predictor of inflation in Italy. Overall, we find that economic news are more relevant in forecasting the monthly variables at short horizons of 1 to 3 months, relative to the case of the GDP where the bulk of the predictability is at intermediate horizons. }

\begin{table}[]
\begin{center}
\setlength{\tabcolsep}{14pt}
%\caption{Unemployment Rate}
\begin{tabular}[t]{lccccc}
\hline\hline\\
Sentiment & DE & ES & FR & IT & UK\\
\hline \\
& \multicolumn{5}{c}{\it Short-run}\\ \\
ECONOMY & 0.971 & 0.998 & {\bf0.027} & {\it 0.082} & {\bf0.029}\\
FINSECTOR & 0.989 & 0.828 & 0.924 & 0.972 & 0.964\\
INFLATION & 0.775 & 0.812 & {\bf0.010} & 0.995 & 0.476\\
MANUFACTURING & 0.730 & {\it 0.096} & 0.280 & 0.992 & 0.223\\
MONETARY POLICY & 0.991 & {\bf0.023} & 0.968 & 0.970 & 0.519\\
UNEMPLOYMENT & {\bf0.043} & {\bf0.007} & 0.965 & 0.977 & 0.895\\ \\
AVERAGE & {\bf0.015} & {\bf0.001} & {\bf0.009} & 0.986 & {\bf0.017}\\
\\
& \multicolumn{5}{c}{\it Long-run}\\ \\
ECONOMY & 0.996 & 0.204 & 0.120 & 0.488 & 0.965\\
FINSECTOR & 0.809 & 0.613 & 0.863 & 0.916 & 0.887\\
INFLATION & 0.327 & {\it 0.056} & {\bf0.013} & 0.981 & 0.922\\
MANUFACTURING & 0.937 & 0.812 & 0.974 & 0.209 & {\it 0.075}\\
MONETARY POLICY & 0.101 & {\bf0.004} & 0.134 & 0.592 & 0.919\\
UNEMPLOYMENT & 0.152 & {\bf0.045} & 0.966 & 0.990 & 0.972\\ \\
AVERAGE & {\bf0.009} & {\bf0.003} & {\bf0.005} & {\bf0.029} & 0.174\\
\\ \hline\hline \\
\end{tabular}
\caption{The dependent variable is \textbf{Unemployment Rate}. Multi-horizon test for equal predictive accuracy when considering exclusively the {\it short-run} horizons (top part), and only the {\it long-run} horizons (bottom part). 
The benchmark model is the ARX and the alternative is the ARXS specification that augments the ARX with a news-based sentiment measure. The Average forecast is obtained from averaging the ARXS forecasts. The value in the table represents the p-value for the one-sided hypothesis that the forecasting models in the column outperform the ARX forecasts: p-values indicating significance at 5\% are denoted in bold and at 10\% in italics.
}
\label{tab:multi_forecast_UNEMP}
% \begin{minipage}{15cm}{Multi-horizon test for equal predictive accuracy when considering exclusively the {\it nowcast} horizons (top part), and only the {\it forecast} horizons (bottom part). 
% The benchmark model is the ARX and the alternative is the ARXS specification that augments the ARX with a news-based sentiment measure. The Average forecast is obtained from averaging the ARXS forecasts. The value in the table represents the p-value for the one-sided hypothesis that the forecasting models in the column outperform the ARX forecasts: p-values indicating significance at least at 10\% are denoted in bold.}
%\end{minipage}
\end{center}
%\label{tab:multi_forecast_UNEMP}
\end{table}

\begin{table}[]
\begin{center}
\setlength{\tabcolsep}{14pt}
%\caption{Industrial Production Index}
\begin{tabular}[t]{lccccc}
\hline\hline\\
Sentiment & DE & ES & FR & IT & UK\\
\hline \\
& \multicolumn{5}{c}{\it Short-run}\\ \\
ECONOMY         & 0.250 & 0.276 & {\it 0.099} & 0.997 & {\bf0.017}\\
FINSECTOR       & 0.856 & 0.855 & 0.117 & 0.998 & 0.571\\
INFLATION       & 0.285 & 0.263 & 0.137 & 0.970 & 0.822\\
MANUFACTURING   & 0.999 & 0.996 & 0.987 & 0.998 & {\bf0.011}\\
MONETARY POLICY & 0.996 &{\bf 0.004} & 0.262 & 0.294 & 0.167\\
UNEMPLOYMENT    & {\it 0.055} & {\bf 0.021} & 0.460 & 0.923 & 0.678\\ \\
AVERAGE         & {\bf0.017} & {\bf 0.012} & {\bf 0.032} & 0.205 & {\bf 0.035}\\ 
\\
& \multicolumn{5}{c}{\it Long-run}\\ \\
ECONOMY         & 0.367 & 0.869 & 0.939 & 0.992 & 0.955\\
FINSECTOR       & 0.916 & 0.987 & 0.981 & 0.980 & 0.897\\
INFLATION       & 0.674 & {\it 0.095} & 0.169 & 0.523 & 0.946\\
MANUFACTURING   & 0.919 & 0.997 & 0.996 & 0.993 & 0.979\\
MONETARY POLICY & 0.945 & {\bf 0.006} & 0.657 & 0.452 & {\bf 0.007}\\
UNEMPLOYMENT    & {\bf0.029} & 0.163 & 0.958 & 0.938 & 0.975\\ \\
AVERAGE         & {\bf 0.025} & {\bf 0.015} & 0.260 & {\bf 0.013} & {\it 0.090}\\
\\ \hline\hline \\
\end{tabular}
\caption{The dependent variable is \textbf{Industrial Production Index}.
Multi-horizon test for equal predictive accuracy when considering exclusively the {\it short-run} horizons (top part), and only the {\it long-run} horizons (bottom part). 
The benchmark model is the ARX and the alternative is the ARXS specification that augments the ARX with a news-based sentiment measure. The Average forecast is obtained from averaging the ARXS forecasts. The value in the table represents the p-value for the one-sided hypothesis that the forecasting models in the column outperform the ARX forecasts: p-values indicating significance at 5\% are denoted in bold and at 10\% in italics.
}
\label{tab:multi_forecast_INDPRO}
% \begin{minipage}{15cm}{Multi-horizon test for equal predictive accuracy when considering exclusively the {\it nowcast} horizons (top part), and only the {\it forecast} horizons (bottom part). 
% The benchmark model is the ARX and the alternative is the ARXS specification that augments the ARX with a news-based sentiment measure. The Average forecast is obtained from averaging the ARXS forecasts. The value in the table represents the p-value for the one-sided hypothesis that the forecasting models in the column outperform the ARX forecasts: p-values indicating significance at least at 10\% are denoted in bold.}
% \end{minipage}
\end{center}
\end{table}

%% CPI
\begin{table}[]
\begin{center}
\setlength{\tabcolsep}{14pt}
%\caption{Consumer Price Index (CPI)}
\begin{tabular}[t]{lccccc}
\hline\hline\\
Sentiment & DE & ES & FR & IT & UK\\
\hline \\
& \multicolumn{5}{c}{\it Short-run}\\ \\
ECONOMY         & {\it 0.099} & 0.988 & 0.989 & 0.123 & 0.975\\
FINSECTOR       & 0.980 & 0.978 & 0.276 & {\bf 0.003} & 0.863\\
INFLATION       & 0.993 & {\it 0.058} & 0.906 & {\it 0.092} & {\bf 0.039}\\
MANUFACTURING   & 0.998 & {\bf 0.009} & 0.983 & 0.992 & 0.675\\
MONETARY POLICY & 0.823 & 0.991 & 0.996 & 0.519 & 0.964\\
UNEMPLOYMENT    & 0.924 & 0.633 & 0.993 & 0.850 & {\it 0.070}\\ \\
AVERAGE         & {\it 0.083} & 0.328 & 0.583 & {\bf 0.004} & {\bf 0.026}\\
\\
& \multicolumn{5}{c}{\it Long-run}\\ \\
ECONOMY         & 0.963 & 0.290 & 0.985 & 0.471 & 0.952\\
FINSECTOR       & 0.987 & 0.602 & 0.130 & {\bf 0.007} & 0.986\\
INFLATION       & 0.997 & 0.168 & 0.987 & 0.625 & 0.948\\
MANUFACTURING   & 0.979 & 0.905 & 0.116 & 0.584 & 0.980\\
MONETARY POLICY & 0.963 & 0.986 & 0.975 & 0.957 & 0.926\\
UNEMPLOYMENT    & 0.826 & 0.227 & 0.988 & 0.988 & {\bf0.024}\\ \\
AVERAGE         & 0.604 & {\bf 0.018} & {\bf 0.044} & {\bf 0.013} & {\bf 0.039}\\ \\ \hline\hline \\
\end{tabular}
\caption{The dependent variable is \textbf{Consumer Price Index}.
Multi-horizon test for equal predictive accuracy when considering exclusively the {\it short-run} horizons (top part), and only the {\it long-run} horizons (bottom part).
The benchmark model is the ARX and the alternative is the ARXS specification that augments the ARX with a news-based sentiment measure. The Average forecast is obtained from averaging the ARXS forecasts. The value in the table represents the p-value for the one-sided hypothesis that the forecasting models in the column outperform the ARX forecasts: p-values indicating significance at 5\% are denoted in bold and at 10\% in italics.
}
\label{tab:multi_forecast_CPI}
% \begin{minipage}{15cm}{Multi-horizon test for equal predictive accuracy when considering exclusively the {\it nowcast} horizons (top part), and only the {\it forecast} horizons (bottom part). 
% The benchmark model is the ARX and the alternative is the ARXS specification that augments the ARX with a news-based sentiment measure. The Average forecast is obtained from averaging the ARXS forecasts. The value in the table represents the p-value for the one-sided hypothesis that the forecasting models in the column outperform the ARX forecasts: p-values indicating significance at least at 10\% are denoted in bold.}
% \end{minipage}
\end{center}
\end{table}

 \newpage

\section{Conclusions\label{sec_conclusion}}

The increased availability of large amounts of textual data, such as news articles, {\color{black} offers the opportunity} to construct text-based sentiment indexes that can potentially complement official macroeconomic and survey-based indicators.
In addition, the Covid-19 pandemic has amplified the need of having timely and reliable economic indicators about different aspects of the current state of economic activity. 
The goal of this paper is to offer an evaluation of the usefulness of sentiment indicators constructed on news articles. We consider a big data set of 27 million articles for the most important newspapers in five European countries, and 
propose a set of \textit{fine-grained}, \textit{aspect-based} sentiment indicators of economic activity. 
Real-time sentiment measures that complement macroeconomic and confidence indicators could be extremely useful for European countries given the considerable delay in the publication of official statistics. 
%{\color{black}\sout{These delays are not negligible: industrial production in Europe has an average publication delay of 30-40 days, relative to 15 days in the US. Furthermore, European statistical agencies do not release any macroeconomic variable at the weekly frequencies that could contribute to an early signal of the direction of economic activity.}}

%% RESULTS
Our findings provide in-sample and out-of-sample evidence that the proposed news-based sentiment indicators produce useful information in forecasting GDP across five European countries. In addition, the predictability provided by the sentiment is incremental relative to the macroeconomic and survey confidence indicators that are available to forecasters. 
We thus conclude that there is a case for considering sentiment measures calculated from news as an additional tool that economic forecasters should adopt in the quest to produce accurate predictions of economic activity. %\sout{We demonstrate this by providing a case study of the application of the indicators to forecasting output during the Covid-19 pandemic. Our findings show that sentiment about the {\it economy} and {\it unemployment} provide real-time information about the softening of economic activity, although the forecasts are not as severe as the decline and rebound of GDP in the second and third quarter of 2020. }
In future work we plan to extend our methodology to perform better in such scenarios. Since our approach is keyword-based, a possible direction could be to extend the set of terms that we consider in constructing the sentiment indicator. On the modelling side, we could make the parameters of the linear forecasting model a function of the number of articles or sentences that discuss the topic in any given day. This would introduce a much needed nonlinearity in the model that could contribute to explain such extreme events.

%Future research could focus on using the proposed sentiment indicators jointly with alternative modeling strategies more suited at forecasting in presence of extreme observations (e.g., see the application of machine learning and regression trees models by \citealp{richardson2020nowcasting} and \citealp{huber2020nowcasting}, respectively) to evaluate the contribution of news-based indicators while forecasting under the pandemic.

%We employ a simple modeling strategy to highlight the overall relevance of the proposed news-based indicators to complement standard macroeconomic information, in particular a forecasting horizons, and their usefulness during the pandemic.
%The added-value of such indicators while nowcasting the Covid-19 outbreak is somewhat limited, probably due to the little flexibility of the consider modeling strategy.

%% FUTURE AGENDA
%Future research could focus on using the proposed sentiment indicators jointly with alternative modeling strategies more suited at forecasting in presence of extreme observations (e.g., see the application of machine learning and regression trees models by \citealp{richardson2020nowcasting} and \citealp{huber2020nowcasting}, respectively) to evaluate the contribution of news-based indicators while forecasting under the pandemic.

%% BIBLIOGRAPHY
\newpage
\spacingset{0.92}
\bibliographystyle{jae} %% Bibliography style Journal of Applied Econometrics
\bibliography{nowcasting}

\end{document}